\pdfoutput=1
\documentclass{llncs}
\usepackage{llncsdoc}
\usepackage{subfigure}
\usepackage{cite}
\usepackage{graphicx}
\usepackage{textcomp}
\usepackage{subfigure}
\usepackage{blindtext}
\usepackage{algorithm,algorithmic}
\usepackage[namelimits]{amsmath} 
\usepackage{amssymb}             
\usepackage{amsfonts}            
\usepackage{mathrsfs}            
\usepackage{booktabs}
\usepackage{multirow}
\usepackage[justification=centering]{caption}
\begin{document}
\markboth{\LaTeXe{} Class for Lecture Notes in Computer
Science}{\LaTeXe{} Class for Lecture Notes in Computer Science}
\thispagestyle{empty}

\vfill
\newpage

\title{Multi-step Cascaded Networks for Brain Tumor Segmentation}

\author{Xiangyu Li \, Gongning Luo \, and  Kuanquan Wang*}
\institute{School of Computer Science and Technology, Harbin Institute of Technology, Harbin 150001, China
}

\maketitle
\begin{abstract}
Automatic brain tumor segmentation method plays an extremely important role in the whole process of brain tumor diagnosis and treatment. In this paper, we propose a multi-step cascaded network which takes the hierarchical topology of the brain tumor substructures into consideration and segments the substructures from coarse to fine .During segmentation, the result of the former step is utilized as the prior information for the next step to guide the finer segmentation process. The whole network is trained in an end-to-end fashion.
Besides, to alleviate the gradient vanishing issue and reduce overfitting, we added several auxiliary outputs as a kind of deep supervision for each step and introduced several data augmentation strategies, respectively, which proved to be quite efficient for brain tumor segmentation. Lastly, focal loss is utilized to solve the problem of remarkably imbalance of the tumor regions and background. Our model is tested on the BraTS 2019 validation dataset, the preliminary results of mean dice coefficients are 0.886, 0.813, 0.771 for the whole tumor, tumor core and enhancing tumor respectively. Code is available at \href{https://github.com/JohnleeHIT/Brats2019}{https://github.com/JohnleeHIT/Brats2019}
\end{abstract}

\section{Introduction}
Brain tumor is one of the most serious brain diseases, among which the malignant gliomas are the most frequent occurred type. The gliomas can be simply divided into two categories according to the severity: the aggressive one (i.e. HGG) with the average life expectancy of nealy 2 years and the moderate one (i.e. LGG) with the life expectancy of several years. Due to the considerably high mortality rate, it is of great importance for the early diagnosis of the gliomas, which largely improves the treatment probabilities especially for the LGG. At present, the most possible ways to treat gliomas are surgery, chemotherapy and radiotherapy. For any of the treatment strategies, accurate imaging and segmentation of the lesion areas are indispensable before and after treatment so as to evaluate the effectiveness of the specific strategy. 

Among all the existing imaging instruments, MRI has been the first choice for brain tumor analysis for its high resolution, high contrast and present no known health threats. In the current clinical routine, manual segmentation of large amount of MRI images is a common practice which turns out to be remarkably time-consuming and prone to make mistakes for the raters. So, it would be of tremendous potential value to propose an automatic segmentation method. Many researchers have proposed several effective methods based on deep learning or machine learning methods to solve the problem. Among those proposed methods, Zikic et al.\cite{zikic2014segmentation}used a shallow CNN network to classify 2D image patches which captured from the MRI data volumes in a sliding window fashion. Zhao et al.\cite{zhao2015deep} converted the 3D tumor segmentation task to 2D segmentation in triplanes and introduced multi-scales by cropping different patch sizes. Havaei et al.\cite{havaei2017brain} proposed a cascaded convolutional network, which can capture local and global information simultaneously.Çiçek et al.\cite{cciccek20163d} extended the traditional 2D U-net segmentation network to a 3D implementation which makes the volume segmentation to a voxel-wise fashion.Kamnitsas et al.\cite{kamnitsas2017efficient} proposed a dual pathway 3D convolution network named DeepMedic to incorporate multi-scale contextual information, and used the 3D fully connected CRF as the postprocess method to refine the segmentation result. Chen et al.\cite{chen2019dual} improved DeepMedic by first cropping 3D patches from multiple layers selected from the original DeepMedic and then merging those patches to learn more information in the network, besides, deep supervision was introduced in the network to better propagate the gradient.Ma.et al\cite{ma2018concatenated}employed a feature representations learning strategy to effectively explore both local and contextual information from multimodal images
for tissue segmentation by using modality specific random forests as the feature learning kernels.

Inspired by Havaei and Çiçek, we proposed a multi-step cascaded network to segment brain tumor substructures. The proposed network uses 3D U-net as the basic segmentation architecture and the whole network works in a coarse-to-fine fashion which can be seen as a kind of spatial attention mechanism.
\section{Methodology}
Based on the thorough analysis of the substructures of brain tumor, which turns out to be a hierarchical topology (see Fig.\ref{fig:structure}), We propose a multi-step cascaded network which is tailored for the brain tumor segmentation task. Our proposed method mainly contains three aspects, detailed information are as follows:

\begin{figure}[ht]
	\vspace{-0.6cm}
	\setlength{\abovecaptionskip}{0.2cm}
	\setlength{\belowcaptionskip}{-0.8cm}
	\centering
	\includegraphics[height=5.2cm,width=8.04cm]{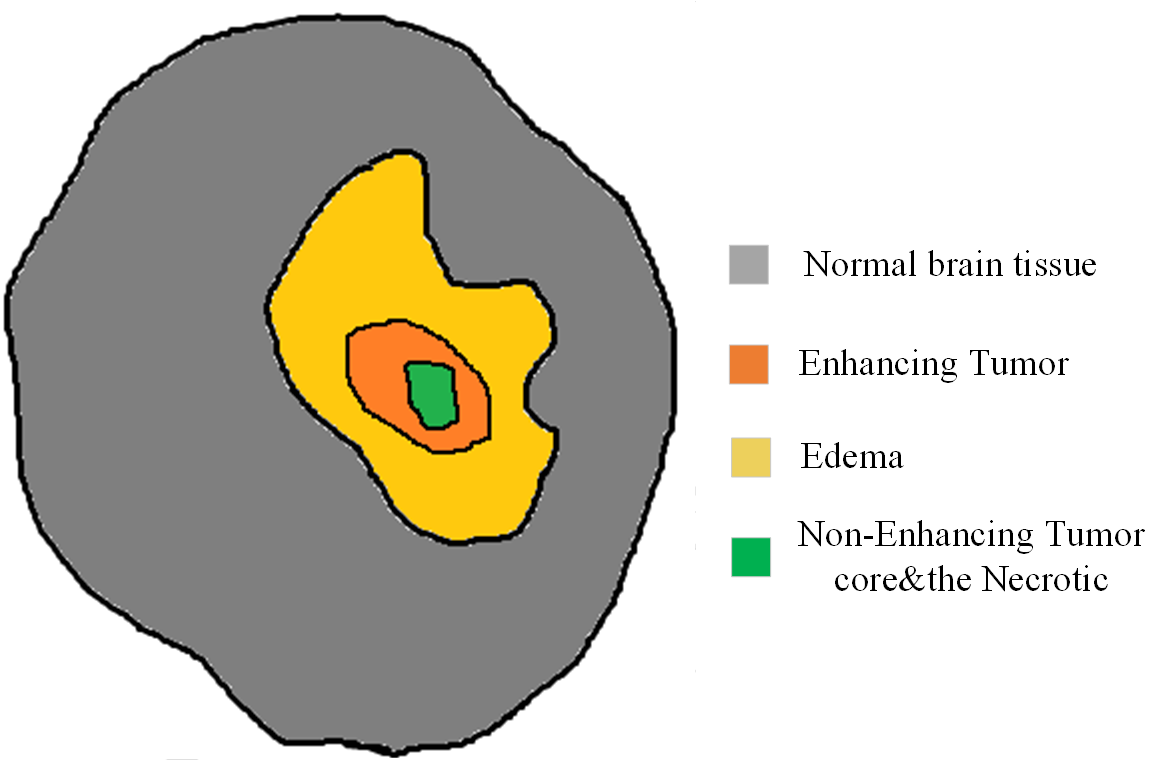}
	\caption[]{ Schematic diagram of the tumor structures}
	\label{fig:structure}
	\vspace{0.5cm}
\end{figure}

\subsection{Multi-step Cascaded Network}
\begin{figure}
	\vspace{-0.3cm}
	\setlength{\abovecaptionskip}{0.2cm}
	\setlength{\belowcaptionskip}{-0.8cm}
	\centering
	\includegraphics[height=5.6cm,width=12.8cm]{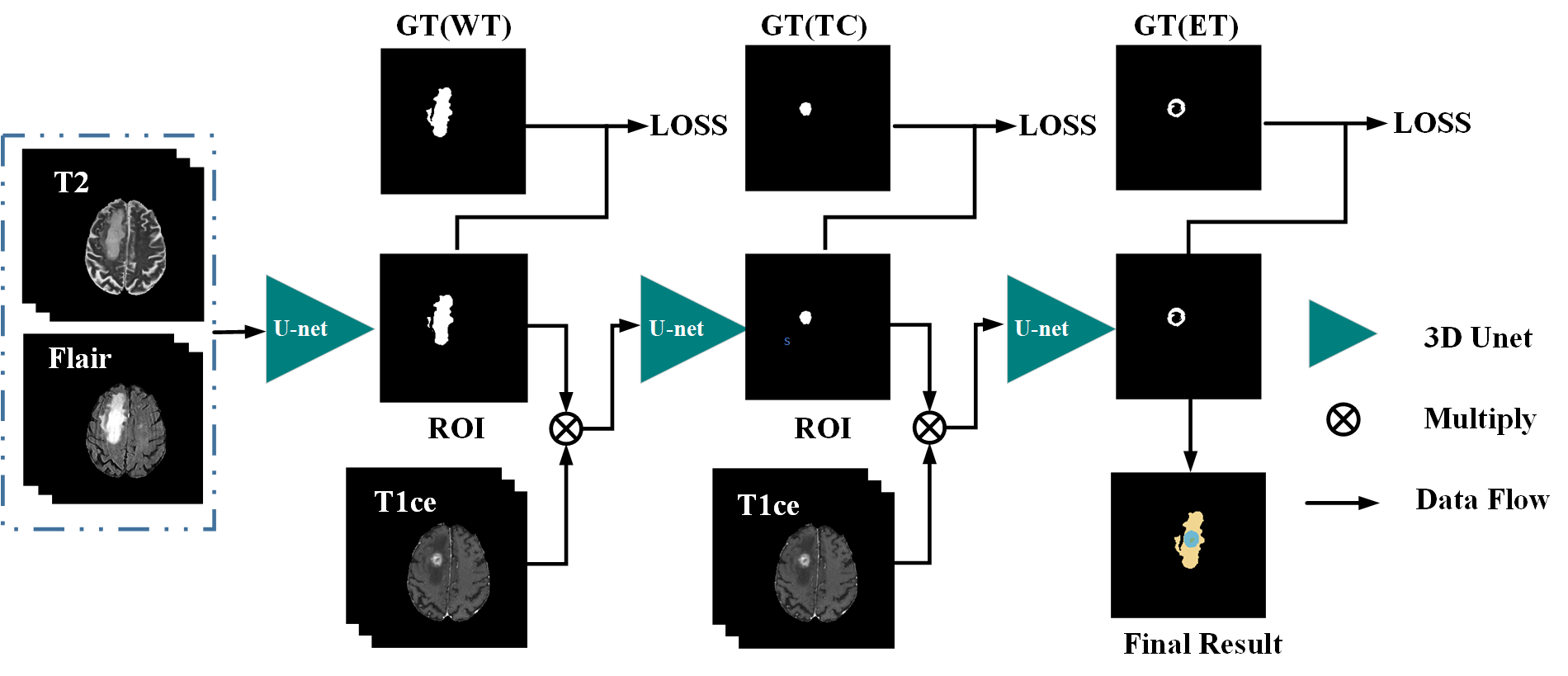}
	\caption[]{ Overview of the proposed multi-step cascaded network}
	\label{fig:network}
\end{figure}

The proposed multi-step cascaded networks are illustrated in Fig.\ref{fig:network}. This method segments the hierarchical structure of the tumor substructures in a coarse-to-fine fashion. In the first step, in order to be consistent with the manual annotations protocol which are detailed descripted in\cite{menze2014multimodal}, two modalities(Flair\&T1ce) of the MRI tumor volumes are utilized.The two-channel data volumes are then fed into the first segmentation network to coarsely segment the whole tumor(WT) which contains all the substructures of the brain tumor; In the second step, similarly, we choose T1ce modality as the data source to segment the tumor core(TC) structure. Besides, the result of the first coarse step can be utilized as the prior information for the second step. By multiplying the mask generated in the first step with the T1ce data volume, the second segmentation network will concentrate more on the corresponding masked areas and make it easier to segment the TC structure. Then the masked volumes are processed by the second network, as a result, TC structure(foreground) are introduced. In the last and finest step, by following the same strategies, we can also get the enhancing tumor (ET) substructures from the data volume, and finally by combining the results of the three steps, the final segmentation maps of the brain tumor will be received. 

\subsection{3D U-net architecture with deep supervisions}

\begin{figure}
\vspace{-0.3cm}
\setlength{\abovecaptionskip}{0.2cm} 
\setlength{\belowcaptionskip}{-0.54cm}
\centering
\includegraphics[height=6.1cm,width=11.5cm]{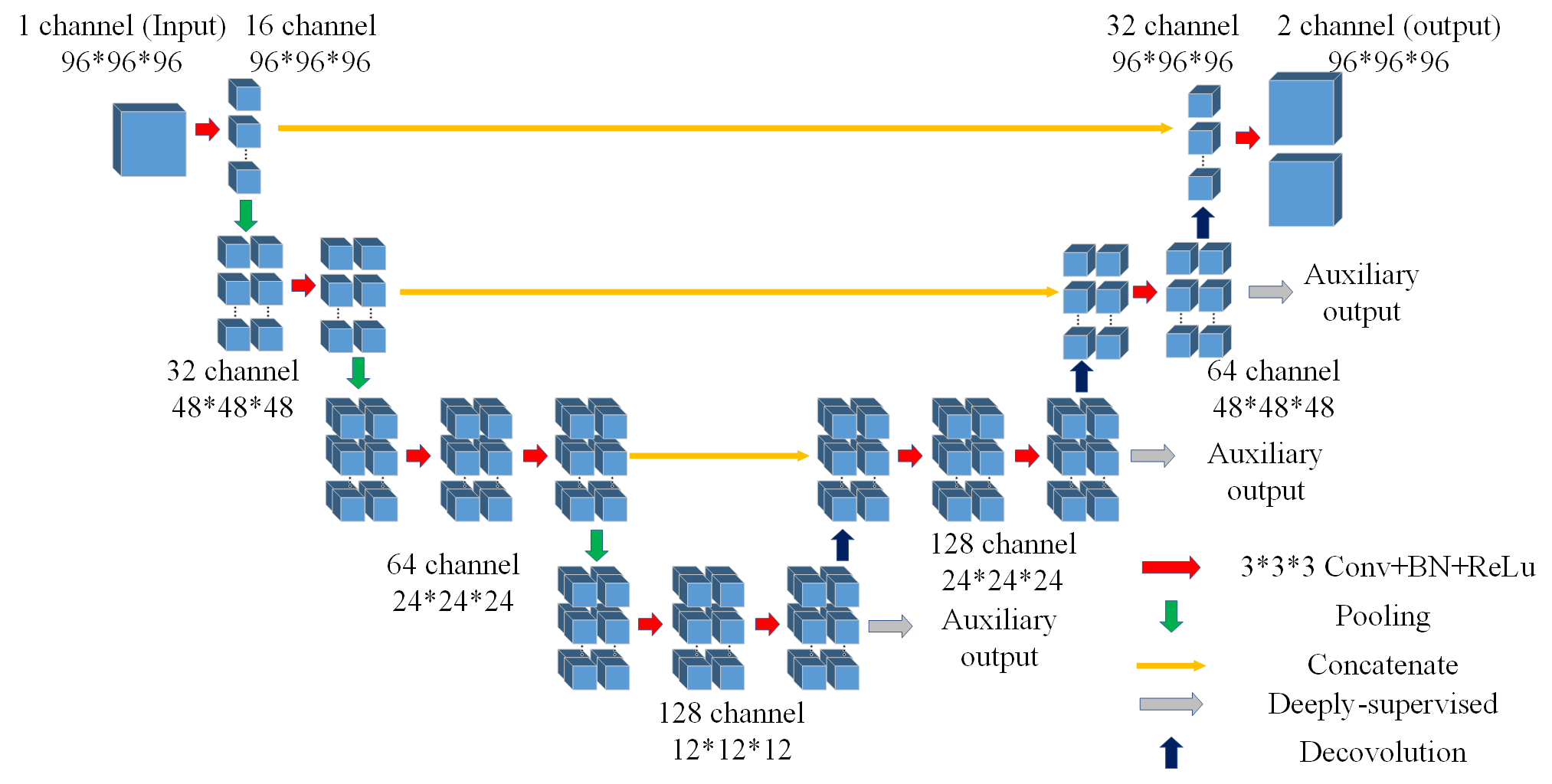}
\caption[]{Schematic of the 3D U-net architecture with deep supervisions} 
\label{fig:unet}
\end{figure}

We take a variant of 3D U-net as the basic segmentation architecture in our multi-step cascaded networks, which is illustrated in Fig.\ref{fig:unet}. The typical 3D U-net consists of two pathways: the contracting pathway and the expanding pathway. The contracting pathway mainly intends to encode the input volumes and introduces the hierarchical features, the expanding pathway however is used to decode the information encoded in the contracting pathway. The two pathways are connected with skip connections so as to make the network be capable of capturing both local and global information. Our basic segmentation network takes 3D U-net as the prototype, whilst makes some improvements on top of it. The main differences between 3D U-net and the proposed basic segmentation networks are as follows:

(1)	Compared to the traditional 3D U-net architecture, our proposed basic segmentation network introduces three auxiliary outputs in the expanding pathway with the intention of better gradient propagation and decreasing the probabilities of vanishing gradient for the relatively deep segmentation networks. As a result, we need to minimize the overall loss functions which comprise both the main branch and the auxiliary loss functions for the basic segmentation process.

(2) We introduce the focal loss\cite{lin2017focal} as the loss function for the whole training process with the intention of alleviating the considerably imbalance of the positive and negative samples in the training data. The focal loss can be expressed as follows:

\begin{equation}
	\mathrm{FL}\left(p_{\mathrm{t}}\right)=-\alpha_{\mathrm{t}}\left(1-p_{\mathrm{t}}\right)^{\gamma} \log \left(p_{\mathrm{t}}\right)
\end{equation}

\begin{equation}
	p_{\mathrm{t}}=\left\{\begin{array}{ll}{p} & {\text { if } y=1} \\ {1-p} & {\text { otherwise }}\end{array}\right.
\end{equation}
where $p \in[0,1]$ is the model's estimated probability for the class with label y=1. $\gamma \geqslant 0$ refers to focusing parameter, it smoothly adjusts the rate at which easy examples are down weighted.$ \alpha_{t}$ refers to balancing factor which balance the importance of the positive and negative samples.

\section {Experiments and Results}
\subsection{preprocessing}

In this paper, we take BraTS 2019 dataset\cite{bakas2017advancing,bakas2018identifying,bakas2017segmentation,bbakas2017segmentation2} as the training data, which comprises 259 HGG and 76 LGG MRI volumes with four modalities (T1, T2, T1ce and Flair) available. According to the official statement of the dataset, all the datasets have been segmented manually following the same annotation protocol. Besides, some preprocessing operations have also been conducted on those datasets, for example, all the MRI volumes have been co-registered to the same anatomical template, interpolated to the same resolution and skull-stripped. Nevertheless, extra preprocessing steps should be done to the raw dataset due to the existence of the intensity nonuniformity in the image data, also called bias field which comes from the imperfect of the MRI machine and the specificity of the patients. This kind of intensity nonuniformity or bias field considerably affects the training process. To eliminate the bias field effect, a great deal of correction methods have been proposed. Among the proposed bias field correction method, the most effective one is the N4 bias field correction\cite{tustison2010n4itk}. In this paper, N4 bias field correction method is utilized as an important preprocessing step before the segmentation process. At last, we also use the normalization method to normalize all the data to zero mean with unit variance.
\subsection{ Implementation Details}
We mixed all the data in the BraTS 2019 training dataset including HGG and LGG, and then trained our model with the mixed dataset. During training, we randomly cropped the raw data volume to sub-volumes with the shape of 96*96*96 due to memory limitation. To reduce overfitting, we introduced some data augmentation methods, for instance, 
rotating a random angle, flipping horizontally and vertically, and adding guassion blur to the sub-volumes with a certain probability. It turned out that the data augmentation was significant important for the brain tumor segmentation task because the network is prone to be overfitting with relatively less training data. We used Adam optimizer to update the weights of the network. The initial learning rate was set to 0.001 at the very beginning and decayed to 0.0005 when the loss curve plateaued. The batch size was set to 1 in the whole training process.

Our model was trained on a Nvidia RTX 2080 Ti GPU for 50 epochs, which takes around 13 hours
\subsection{ Segmentation Results}
To evaluate our proposed mothed, we tested our algorithm on both training and validation set by uploading the inference results to the online evaluation platform (CBICB’s IPP), we finally got the evaluation results including Dice sore, Hausdorff distance, sensitivity and specificity for the whole tumor(WT), the tumor core(TC) and the enhancing tumor(ET), respectively. The metrics aforementioned are defined as follows:

\begin{equation}
	\operatorname{Dice}(P, T)=\frac{\left|P_{1} \wedge T_{1}\right|}{\left(\left|P_{1}\right|+\left|T_{1}\right|\right) / 2}
\end{equation}

\begin{equation}
	\operatorname{Sensitivity}(P, T)=\frac{\left|P_{1} \wedge T_{1}\right|}{\left|T_{1}\right|}
\end{equation}

\begin{equation}
	\operatorname{Specificity}(P, T)=\frac{\left|P_{0} \wedge T_{0}\right|}{\left|T_{0}\right|}
\end{equation}

\begin{equation}
	\operatorname{Haus}(T, P)=\max \left\{\operatorname{supinf}_{t \in T p \in P} d(t, p), \operatorname{supin}_{p \in P} \inf d(t, p)\right\}
\end{equation}
where $P$ refers to the prediction map of the algorithm, and $T$ is the groundtruth label segmented manually by the experts. $\wedge$ is the logical AND operator, $|\cdot|$ means the number of voxels in the set, and $P_{1}$,$P_{0}$ represent the postive and negative voxels in the prediction map, respectively, and $T_{1}$,$T_{0}$ denote the positive and negative voxels in the groundtruth map, respectively. 

Table.\ref{tab:Tab03} presents the quantitative average results on both training and validation dataset. Not surprisingly, the dice coefficient and sensitivity of the whole tumor, the tumor core and the enhancing tumor are in a descending order for both datasets due to the ascending difficulties for those tasks. However, there still exists small gaps for the evalutation metrics between training and validation dataset which attributed to the overfitting problem.
\begin{table*}[]
	\vspace{-0.8cm}	
	\scriptsize	
	\centering	
	\caption{Quantitative average results on the training and validation dataset}	
	\label{tab:Tab03}
	\setlength{\tabcolsep}{3.5mm} 
	\begin{tabular}{cccccc}	
		\toprule
		Dataset &Label &Dice & Sensitivity & specificity & Hausdorff Distance \\				
		\midrule
		\multirow{3}{*}{Training} &WT &0.915 &\textbf{0.942} &0.993 &4.914 \\
		&TC &0.832 &0.876 &0.996 &6.469\\
		&ET &0.791 &0.870 &0.997 &6.036 \\
		\multirow{3}{*}{Validation} &WT &0.886 &\textbf{0.921} &0.992 &6.232 \\
		&TC &0.813 &\textbf{0.819} &0.997 &7.409 \\
		&ET &0.771 &\textbf{0.802} &0.998 &6.033 \\
		\bottomrule	
	\end{tabular}	
	\setlength{\abovecaptionskip}{-2cm}
	\setlength{\belowcaptionskip}{-5.4cm}
	\vspace{-0.5cm}
\end{table*}

\begin{figure}[ht]
	\vspace{-0.9cm}
	\setlength{\abovecaptionskip}{0.2cm} 
	\setlength{\belowcaptionskip}{0.26cm}
	\centering
	\includegraphics[height=9.2cm,width=12.57cm]{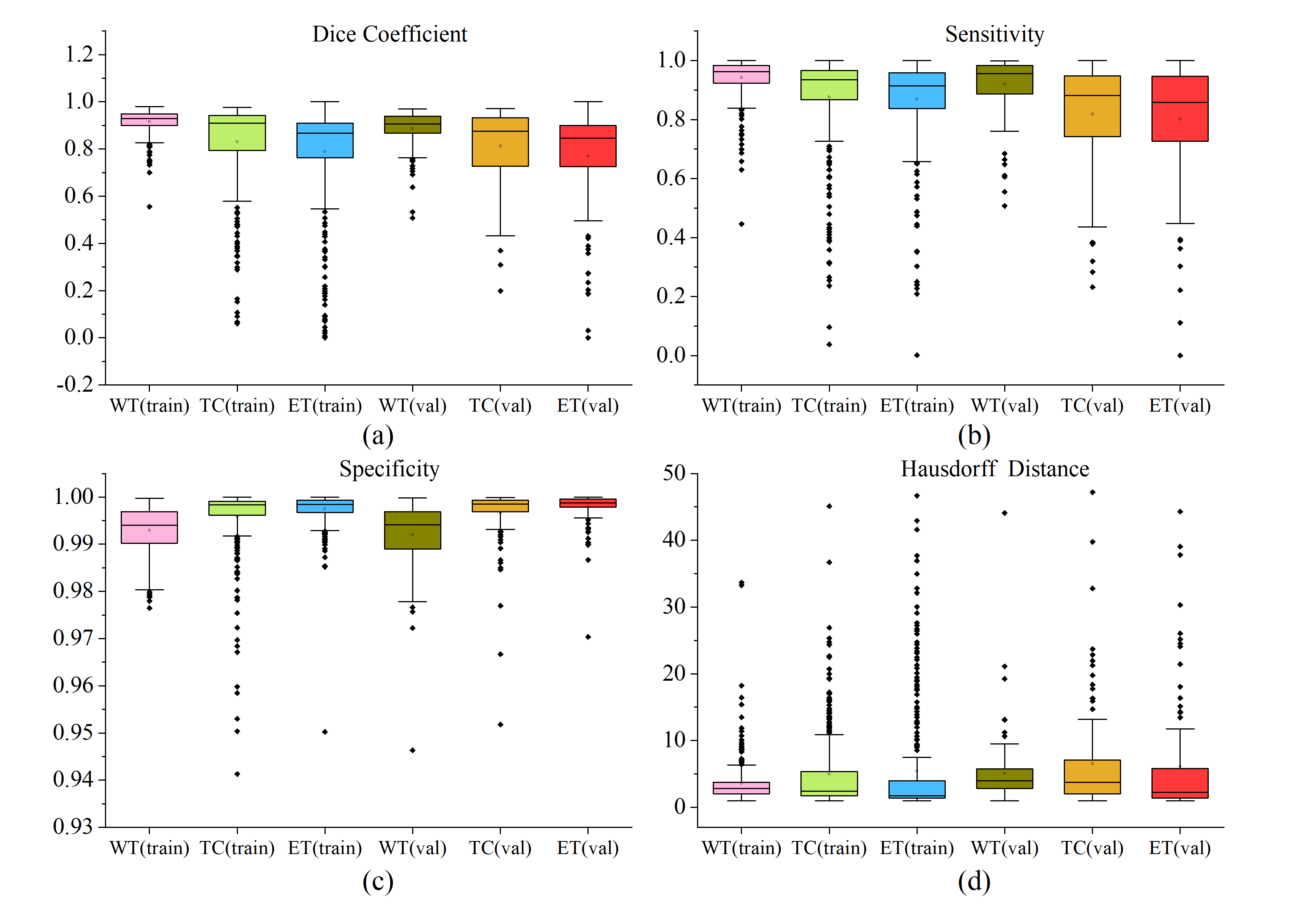}
	\caption[]{Boxplot of the overall performance on both training and validation datasets} 
	\label{fig:validation}
	\vspace{-0.9cm}
\end{figure}

To better analysis the overall performance of the proposed algorithm, we made the boxplot of all the validation and training results, which can be seen from Fig.\ref{fig:validation} . It is evident that the proposed method can segment well on almost all the volumes in both datasets except for a few outliers. Besides, by comparing the boxplot of the validation and training dataset, we noticed that the variance of all the evaluation metrics including dice coefficient,sensitivity,specificity and hausdorff distance for the validation dataset is larger than those for training dataset, which means that our method still suffers from the overfitting problem to some extent. Finally, we can see from the 4 subgraphs that the variance of dice coefficient for the whole tumor is smaller than both tumor core and enhancing tumor substructures for both training and validation datasets, the same for sensitivity and hausdorff distance metrics and the opposite for the specificity metrics, which are in line with our expectations.However, what surprise us most is that the variance of the tumor core(TC) is larger than that of enhancing tumor core(ET) on most metrics for the two datasets, the most possible explanation of the fact is that the network sometimes predicts the whole tumor as the tumor core mistakenly with the impact of the LGG tumor samples, which increases the variations sharply.

\begin{figure}
	\vspace{-0.05cm}
	\setlength{\abovecaptionskip}{0.2cm} 
	\setlength{\belowcaptionskip}{-0.54cm}
	\centering
	\includegraphics[height=12cm,width=11.5cm]{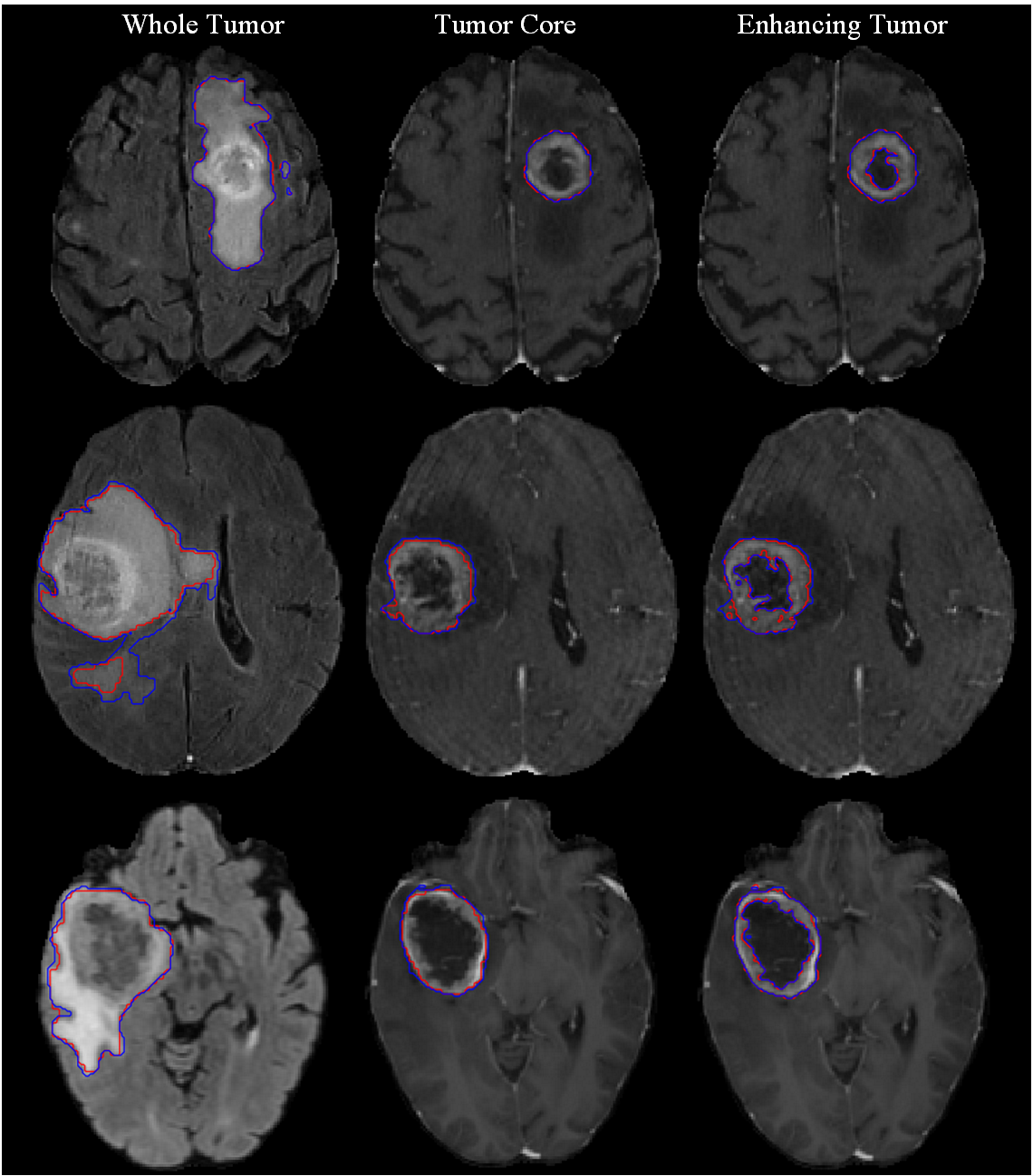}
	\caption[]{Segmentation result of the whole tumor (WT), Tumor core (TC) and Enhancing tumor (ET) structures for HGG tumors, each shows the ground truth label (The blue line) and the prediction result (The red line)} 
	\label{fig:HGG_result}
\end{figure}

\begin{figure}
	\vspace{-0.45cm}
	\setlength{\abovecaptionskip}{0.2cm} 
	\setlength{\belowcaptionskip}{-0.54cm}
	\centering
	\includegraphics[height=12cm,width=11.5cm]{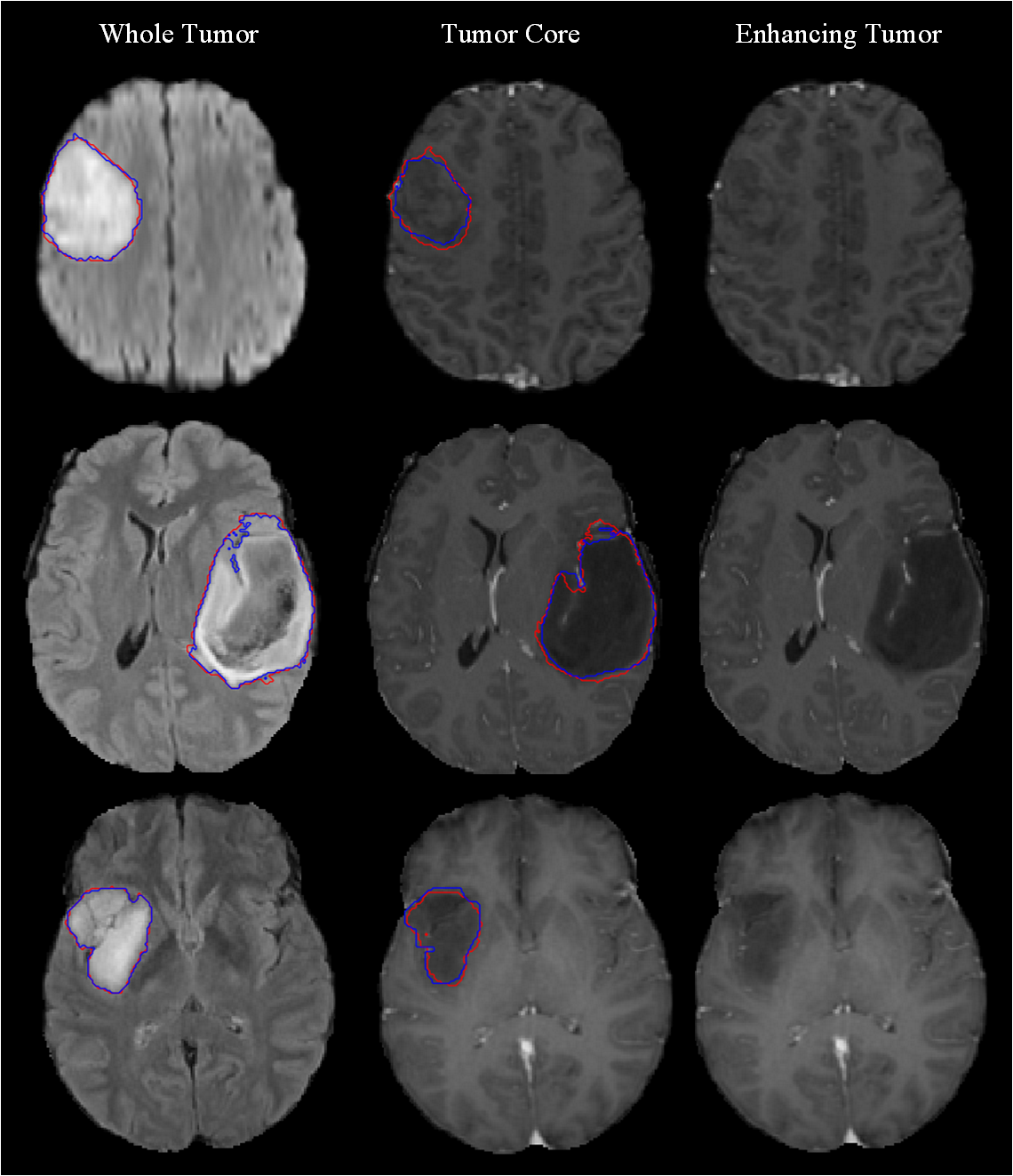}
	\caption[]{Segmentation result of the whole tumor (WT), Tumor core (TC) and Enhancing tumor (ET) structures for LGG tumors, each shows the ground truth label (The blue line) and the prediction result (The red line)} 
	\label{fig:LGG_result}
\end{figure}

Qualitative analysis of the segmentation results for the HGG and LGG tumors are also introduced, which can be seen from Fig.\ref{fig:HGG_result} and Fig.\ref{fig:LGG_result}, respectively. The left row are the flair modality images with the whole tumor ground truth and the prediction result, demonstrated in blue and red curves respectively. The middle row are the T1ce modality images with the tumor core ground truth and the prediction result which are illustrated in the same way as the left row. The right row of course focus on the remaining substructure, i.e. the enhancing tumor. 

All of the three regions with great clinical concerns have been well segmented except for some small details. Not surprisingly, our aforementioned guess about the difficulties of the three tasks can be verified again from the visualization result. Specifically, from step one to step three, the task becomes tougher because the contrast between the tumor region and the surrounding background decreases and the segmentation substructures contours become much rougher at the same time. 
\section {Discussing and Conclusion}
By visualizing all the validation results, we find it interesting that plenty of bad segmented cases for the tumor core regions are those who mistaken the whole tumor as the tumor core region. The most possible explanation might be the variations between different MRI volumes despite the same modality. So, it is likely that the results would increase if some preprocessing methods which can decrease those variations have been taken before the training process, e.g. histogram equalization.
 Besides, we also tried the curriculum learning strategy which trained the network step by step instead of end-to-end training, it turns out that the results are no better than the end-to-end training ones. That is most likely because the network can fit the training data better if all the parameters in the network can be updated. Lastly, we tried to weight the three steps of the cascaded network, surprisingly, we find that the final results present no big difference for increment, decrement or even weights of the training steps.

In conclusion, we present a very efficient multi-step network to segment all the tumor substructures. We first choose specific modalities for each step to keep the automatic segmentation process to be consistent with the mamual protocol which improves our result a lot compared to the method to use all the modalities. After that, we preprocess the input volumes with N4 bias field correction and normalization. Due to the memory limitation, we randomly crop volume patches from the original data and introduce data augmentation on those patches, We find the data augmentation is quite important for reducing overfitting especially when the training data is scarce.  

At last, the training patches are trained in the multi-step network which has proved to be more effective than the one-step couterpart as it trains the network in a coarse-to-fine fashion and seperates the tough multi-classification problem to three much easier binary-classification issuse.

We evaluated the proposed mothod on the BraTS 2019 validation dataset, the results show that our method performance well on all three substructures.

\section {Acknowledgments}
This work was supported by the National Key R\&D Program of China under Grant 2017YFC0113000

\bibliographystyle{splncs}
\bibliography{tumor_reference}
\end{document}